# Interactions between Polymers and Nanoparticles : Formation of « Supermicellar » Hybrid Aggregates

**J.-F. Berret[@], K. Yokota\* and M. Morvan,**
Complex Fluids Laboratory, UMR CNRS - Rhodia n°166,
Cranbury Research Center Rhodia 259 Prospect Plains Road
Cranbury NJ 08512 USA

**Abstract :**

When polyelectrolyte-neutral block copolymers are mixed in solutions to oppositely charged species (e.g. surfactant micelles, macromolecules, proteins etc…), there is the formation of stable "supermicellar" aggregates combining both components. The resulting colloidal complexes exhibit a core-shell structure and the mechanism yielding to their formation is electrostatic self-assembly. In this contribution, we report on the structural properties of "supermicellar" aggregates made from yttrium-based inorganic nanoparticles (radius 2 nm) and polyelectrolyte-neutral block copolymers in aqueous solutions. The yttrium hydroxyacetate particles were chosen as a model system for inorganic colloids, and also for their use in industrial applications as precursors for ceramic and opto-electronic materials. The copolymers placed under scrutiny are the water soluble and asymmetric poly(sodium acrylate)-*b*-poly(acrylamide) diblocks. Using static and dynamical light scattering experiments, we demonstrate the analogy between surfactant micelles and nanoparticles in the complexation phenomenon with oppositely charged polymers. We also determine the sizes and the aggregation numbers of the hybrid organic-inorganic complexes. Several additional properties are discussed, such as the remarkable stability of the hybrid aggregates and the dependence of their sizes on the mixing conditions.

* Present address : Rhodia, Centre de Recherches d'Aubervilliers, 52 rue de la Haie Coq, F-93308 Aubervilliers Cedex France

# I - Introduction

In the phosphor, superconductor and ceramic manufacturing, rare earth oxides and yttrium oxide in particular are of great importance, and thus they have found multiple industrial applications [1,2]. In recent years, particles made from these oxides and with sizes down to the nanometer range (1 − 100 nm) have attracted lots of interest. The use of ultrafine particles not only favors a good dispersibility or a very highly densification of the final products [3] but also improve their performances. In luminescent displays for instance, ultrafine particles are a crucial ingredient because they allow the design of high resolution panels [4,5]. For applications, the traditional process to obtain fine rare earth oxide particles (or precursors) is the precipitation of rare earth salts by alkaline solutions, followed by the separation of the aggregates, the drying and calcination, eventually accompanied by mechanical milling. With this method, particles remain large, at a few hundreds nanometers in size. Moreover, their size distribution is not properly controlled. A number of papers have proposed alternative methods [6,7], using mostly organic solvents such as in the sol-gel or in the microemulsion processing. In industrial plants however, due to intrinsic economical and environmental charges associated to the use of organic solvents, strict limitations are now being applied. The same trend is true for the manufacturing of the end-products. Therefore, the aqueous processing is a key feature for the future applications of ultrafine rare earth particles [8,9].

Using a two-step synthesis, we have developed recently ultra-fine yttrium hydroxyacetate nanoparticles in aqueous solutions that could fulfill the requirements for finding new materials [10,11]. The yttrium hydroxyacetate particles are spherical in average (radius 2 nm), rather monodisperse and they are positively charged at neutral pH. They are regarded as a precursor of yttrium oxide. For colloidal dispersions in general, the issue of the stability is crucial. The stability of the yttrium precursors at neutral pH is excellent over a period of several weeks, after which a slow and progressive precipitation is observed. The origin of this precipitation is not yet understood. In order to improve the stability of the yttrium hydroxyacetate nanoparticles and then prevent precipitation, we have studied the complexation of the precursor nanoparticles with oppositely charged block copolymers. These copolymers placed under scrutiny are the water soluble poly(sodium acrylate)-*b*-poly(acrylamide) diblocks, noted in the sequel of the paper PANa-*b*-PAM. In PANa-*b*-PAM, the poly(sodium acrylate) block is electrostatically charged and of opposite charge to that of the particles and the poly(acrylamide) is neutral. When polyelectrolyte-neutral block copolymers are mixed in solutions to oppositely charged species (e.g. surfactant micelles, macromolecules, proteins etc…), there is the formation of stable "supermicellar" aggregates combining both components [12-30]. The resulting colloidal complexes exhibit a core-shell structure and the mechanism yielding to their formation is known as electrostatic self-assembly [14,31]. A schematic representation of the colloidal complexes made from diblocks and surfactants [22-24,28,29] is displayed in Fig. 1. In this work, using dynamical and static light scattering, we demonstrate the formation of colloidal complexes resulting from the spontaneous association of inorganic ultra-fine yttrium particles and block copolymers. The complexation technique enables to control the aggregation of the organic-inorganic hybrid colloids in the range 20 - 50 nm.

# II - Experimental

## II. 1 - Characterization and Sample Preparation

### II.1.1 - Polyelectrolyte-Neutral Diblock Copolymer

In this work, surfactant micelles and yttrium-based nanoparticles have been complexed with the oppositely charged poly(acrylic acid)-*b*-poly(acrylamide) block copolymers (Fig. 2). The synthesis of this polymer is achieved by controlled radical polymerization in solution [32,33]. Poly(acrylic acid) is a weak polyelectrolyte and its ionization degree (*i.e.* its charge) depends on the pH. All the experiments were conducted at neutral pH (pH 7) where ~ 70 % of monomers are negatively charged. In order to investigate the role of the chain lengths on the formation of complexes, copolymers with different molecular weights were synthesized [29]. In the present study, we focus on a unique diblock, noted PANa(69)-*b*-PAM(840). The two numbers in parenthesis are the degrees of polymerization of each block. These numbers correspond to molecular weights 5000 and 60000 g mol$^{-1}$, respectively. The abbreviation PANa stands here for poly(sodium acrylate), which is the sodium salt of the polyacid. Static and dynamic light scattering experiments were conducted for the determination of the weight-average molecular weight $M_w^{pol}$ and the mean hydrodynamic radius $R_H^{pol}$ of the chains. In water, PANa(69)-*b*-PAM(840) is soluble and its coil configuration is associated to an hydrodynamic radius $R_H^{pol}$ = 8 nm. The weight-average molecular weight is 68 300 ± 2 000 g mol$^{-1}$, in good agreement with the values anticipated from the synthesis. The polydispersity index of the polymers was estimated by size exclusion chromatography at 1.6.

### II.1.2 - Inorganic Nanoparticles

As already mentioned, we are concerned here with the formation of complexes between PANa(69)-*b*-PAM(840) copolymers and yttrium nanoparticles. In order to establish a link with previous studies, we will recall some data obtained on oppositely charged surfactants such as the dodecyltrimethylammonium bromide (DTAB) [22-24,28,29]. DTAB is a cationic surfactant with 12 carbon atoms in the aliphatic chain. it was purchased from Sigma and used without further purification. Its critical micellar concentration is 0.46 wt. % (15 mmol l⁻¹).

The synthesis of the yttrium hydroxyacetate nanoparticles is based on two chemical reactions. One is the dissolution of high purity yttrium oxide by acetic acid and the second is a controlled reprecipitation of yttrium hydroxyacetate obtained on cooling, yielding for the average composition of the particles $Y(OH)_{1.7}(CH_3COO)_{1.3}$. We have determined by light scattering the molecular weight ($M_w^{nano}$ = 27 000 g mol⁻¹) and the hydrodynamic radius of these nanoparticles ($R_H^{nano} \sim 2$ nm). Zeta potential measurements confirm that they are positively charged ($\zeta = +45$ mV) and that the suspensions are stabilized by surface charges. The stability of $Y(OH)_{1.7}(CH_3COO)_{1.3}$ sols at neutral pH is excellent over a period of several weeks. The radius of gyration of the particles was finally determined by small-angle neutron and x-ray scattering. It was found respectively at $R_G^{nano} = 1.82$ nm and 1.65 nm, again in good agreement with the dynamical light scattering result.

*II.1.3 - Sample Preparation*

The supermicellar aggregates were obtained by mixing a surfactant or a nanoparticle solution to a polymer solution, both prepared at the same concentration c and same pH. For the surfactant-polymer complexes, the relative amount of each component is monitored by the charge ratio Z. For the DTAB surfactant and PANa(69)-*b*-PAM(840) copolymers, Z is given by [DTAB]/(69×[PANa(69)-*b*-PAM(840)]) where the quantities in the square brackets are the molar concentrations for the surfactant and for the polymer. Z = 1 describes a solution characterized by the same number densities of positive and negative chargeable ions. Polymer-nanoparticles complexes were prepared using similar protocols [11]. For this system, the charge ratio could not be used since the average structural charge borne by the particles is not known. Here, only the effective charge is known [10] and it is of the order of $4R_H^{nano}/\ell_B \sim 11$, where $\ell_B = 0.7$ nm is the Bjerrum length in water [34]. We define instead a mixing ratio X which is the volume of yttrium-based sol relative to that of the polymer. According to the above definitions, the concentrations in nanoparticles and polymers in the mixed solutions are $c_{pol} = c/(1+X)$ and $c_{nano} = Xc/(1+X)$.

## II. 2 - Light Scattering Techniques and Data Analysis

Static and dynamic light scattering were performed on a Brookhaven spectrometer (BI-9000AT autocorrelator) for measurements of the Rayleigh ratio $\mathcal{R}(q,\mathbf{c})$ and of the collective diffusion constant $D(\mathbf{c})$. A Lexel continuous wave ionized Argon laser was operated at low incident power (20 mW - 150 mW) and at the wavelength $\lambda = 488$ nm. The wave-vector q is defined as $q = \dfrac{4\pi n}{\lambda}\sin(\theta/2)$ where n is the refractive index of the solution and $\theta$ the scattering angle. Light scattering was used to determine the apparent molecular weight $M_{w,\text{app}}$ and the radius of the gyration $R_G$ of the supermicellar aggregates. In the regime of weak colloidal interactions, the Rayleigh ratio follows the classical expression for macromolecules and colloids [35,36] :

$$\frac{K\,c}{\mathcal{R}(q,c)} = \frac{1}{M_{w,\text{app}}}\left(1 + \frac{q^2 R_G^2}{3}\right) + 2A_2 c \qquad (1)$$

In Eq. 1, $K = 4\pi^2 n^2 (dn/dc)^2 / \mathcal{N}_A \lambda^4$ is the scattering contrast coefficient ($\mathcal{N}_A$ is the Avogadro number) and $A_2$ is the second virial coefficient. The refractive index increments dn/dc were measured on a Chromatix KMX-16 differential refractometer at room temperature. The values of the refractive index increments for the different solutions are reported in Table I. With light scattering operating in dynamical mode, we have measured the collective diffusion coefficient D($\mathbf{c}$) in the range c = 0.01 wt. % − 1 wt. %. From the value of D(c) extrapolated at c = 0, the hydrodynamic radius of the colloids was calculated according to the Stokes-Einstein relation, $R_H = k_B T / 6\pi\eta_0 D_0$, where $k_B$ is the Boltzmann constant, T the temperature (T = 298 K) and $\eta_0$ the solvent viscosity ($\eta_0 = 0.89 \times 10^{-3}$ Pa s). The autocorrelation functions of the scattered light were interpreted using the method of cumulants. For the supermicellar aggregates, the diffusion coefficients derived by this technique appeared to be identical above linear term (i.e. the quadratic, cubic and quartic).

# III − Results and Discussion

## III. 1 - The Surfactant-Nanoparticle Analogy

In Figs. 3 are displayed the scattering properties of mixed solutions of block copolymers and oppositely charged surfactants [29]. These properties are the Rayleigh ratio $\mathcal{R}(q,c)$ at $\theta$ = 90° and the hydrodynamic radius $R_H$. These Properties are shown as function of the charge ratio Z. The solutions were prepared as described in the experimental section by mixing a polymer solution at c = 1 wt. % to the surfactant solution at the same concentration. Hence, each data points in Figs. 3a and 3b corresponds to a distinct solution.

At low Z (Z < 1), the scattering intensity is independent of Z and it remains at the level of the pure polymer. The hydrodynamic radii $R_H$ are also close to those of single chains. With increasing Z, there is a critical charge ratio $Z_C$ above which the Rayleigh ratio increases noticeably. $\mathcal{R}(q,\mathbf{c})$ then levels off in the range Z = 1 – 10, and decreases at higher Z values. Dynamical light scattering performed above $Z_C$ reveals the presence of a purely diffusive relaxation mode, associated to Brownian diffusion. For the PANa(69)-*b*-PAM(840)/DTAB, the hydrodynamic radius $R_H$ ranges between 50 nm and 70 nm. Such hydrodynamic values are well above those of the individual components, a result that suggests the formation of mixed aggregates. The microstructure of such aggregates (Fig. 1) has been discussed at length in previous reports and we refer to them for a quantitative description [22-24,28,29]. Because this structure is finally very similar to that of polymeric micelles made from amphiphilic diblocks [14,17,18], the mixed surfactant-polymer complexes are sometimes referred to as a "supermicelles" or a "supermicellar" aggregates.

We have repeated the same experiments with nanoparticles and polymers. The yttrium hydroxyacetate sol was mixed with a c = 1 wt. % PANa(69)-*b*-PAM(840) solution at different volumic ratios. Figs. 4a and 4b display the $\mathcal{R}(q,c)$ measured at $\theta = 90°$ and the hydrodynamic radius $R_H$. The experimental conditions are the same as those of Fig. 3. The results between surfactant-polymers and nanoparticle-polymer mixtures are qualitatively similar. Above a critical value of the mixing ratio noted $X_C$ ($X_C \sim 0.1$), the scattering increases and the hydrodynamic radius saturates at values comprised between 30 nm and 40 nm. As for the surfactant system, dynamical light scattering reveals the formation of rather monodisperse colloids of hydrodynamic sizes much larger than the individual components. In electrostatic self-assembly, there is classically a mixing ratio, noted $X_P$ at which all the components present in the solution react and form complexes [14,30,31]. $X_P$ corresponds roughly to a state where the cationic and anionic charges are in equal amounts. Experimentally, it is the ratio where the number density of complexes is the largest, *i.e.* where the scattering intensity as function of X presents a maximum. From the X-dependence of the Rayleigh ratio (Fig. 4a), we find for the yttrium–diblock system $X_P$ = 0.2. Using the molecular weights of the single components (Table I) and the relationship $X_P = \overline{n^{nano}}M_w^{nano} / \overline{n^{pol}}M_w^{pol}$, the value of $X_P$ allows us to find the number of polymers per particle involved in the complexation. One gets :

$$\overline{n^{pol}} / \overline{n^{nano}} \sim 2 \qquad (2)$$

In the sequel of the paper, we assume that the hybrid nanoparticle-polymer complexes, when they form, are at the preferred composition $X_P$. $\overline{n^{pol}}$ and $\overline{n^{nano}}$ are the average aggregation numbers that characterize the mixed colloids.

Above $X_P$, the intensity decreases with increasing X, and an explanation for this behavior can be found again in the complexation mechanism. At large X, the nanoparticles are in excess with respect to the polymers, and thus all the polymers are consumed to build the supermicellar colloids. Since the polymer concentration decreases with X according to $c_{pol}$ = c/(1+X), the number of these colloids and thus the scattering intensity decreases. Assuming furthermore $\overline{n^{pol}}$ to be a constant *versus* X, the intensity should explicitly decrease as 1/X. As shown in Fig. 4b by a straight line in the double logarithmic plot, this behavior is indeed observed for the yttrium-polymer system. At X > 10, we find actually a state of coexistence between the mixed nanoparticle-polymer aggregates and the uncomplexed nanoparticles at $R_H \sim 2$ nm. By comparing Figs. 3 and Figs. 4, we demonstrate finally the analogy between surfactant micelles and nanoparticles in the complexation phenomenon with oppositely charged polymers.

## III.2 – Hydrodynamic Sizes for the nanoparticle-polymer complexes

In order to determine more accurately the sizes of the colloidal complexes, we have performed dynamical light scattering as function of the concentration. In Fig. 5 the quadratic diffusion coefficient D(c) is shown *versus* c for three series of samples. These solutions were obtained by dilution of the samples prepared at c = 1 wt. % (X = 0.2, empty circles; X = 0.5, triangles) and at c = 4 wt. % (X = 0.25, closed circles). In this concentration range, the diffusion coefficient varies according to :

$$D(c) = D_0(1 + D_2c) \qquad (3)$$

where $D_0$ is the self-diffusion coefficient and $D_2$ is a virial coefficient of the series expansion [37]. From the sign of the virial coefficient, the type of interactions between the aggregates, either repulsive or attractive can be deduced. Here $D_2$ is small (of the order of $10^{-7}$ $cm^2s^{-2}$), indicating that the interactions between colloids are weak in this concentration range. From the values of the self-diffusion coefficient $D_0$, $R_H$ is derived and it is found in the range 28 nm - 42 nm (see Table I for details). The data in Fig. 5 shows finally two important properties for the nanoparticle-polymer complexes.

1 - Once the supermicellar aggregates are formed, they remain stable upon dilution down to concentrations of the order of 0.01 wt. %. This result indicates that the critical aggregation concentration (cac) for the self-assembled colloids is certainly below this value. Working at concentrations lower than 0.01 wt. % in light scattering becomes difficult because of the low level of the scattering signal.

2 - The second important result found in Fig. 5 is that hydrodynamic sizes of the complexes depend slightly on the mixing conditions. For instance, mixing at high concentrations yields larger colloids. Although this observation is rather general (it was also observed with surfactants [23]), it is not yet fully understood.

## III.3 - Aggregation numbers

In this section, we determine the average aggregation numbers for the nanoparticle-polymer complexes. We assume that each mixed solution at $X > X_C$ is a dispersion of nanoparticle-polymer aggregates and that these aggregates are characterized by the aggregation numbers $\overline{n^{nano}}$ and $\overline{n^{pol}}$. To this aim, we focus on the series of solutions that have been prepared at the preferred composition $X_P$, or close to it. Figs. 6 and 7 use the Zimm representation [35] to display the light scattering intensities for these samples. The solutions are those of Fig. 5 at $X = 0.2$ (Fig. 6) and $X = 0.25$ (Fig. 7). Here, the quantity $Kc/\mathcal{R}(q,\mathbf{c})$ is plotted as function of $q^2 + cste \times c$ and it is compared to the predictions of Eq. 1. These predictions, shown as straight lines in the two figures allow us to determine the radius of gyration $R_G$ and the molecular weight $M_{w,app}$ for the aggregates. Their values are reported in Table I. In both cases, the virial coefficient $A_2$ is weak and of the order of $10^{-6}$ $cm^{-3}$ $g^{-2}$. In Table I, we also include the results for the $X = 0.5$-series. The radius of gyration ranges between 20 and 30 nm, and the ratio $R_G/R_H$ is found around $0.6 - 0.7$. It is interesting to note here that this later values are characteristic for core-shell structures. They are typical for instance for polymeric micelles composed from block copolymers in selective solvent [38-40].

The apparent molecular weight of the complexes are of the order of $1 - 10 \times 10^6$ g mol$^{-1}$. For colloids resulting from a self-assembly process, the apparent molecular weight can be expressed as [36]:

$$M_{w,\text{app}} = \left( \overline{n^2} / \overline{n} \right) m_n^* + \left( m_w^* - m_n^* \right) \tag{4}$$

where $\overline{n}$ and $\overline{n^2}$ are the first and second moments of the distribution of aggregation numbers. In Eq. 4, $m_n^*$ and $m_w^*$ are the number and weight-average molecular weights of the elementary building blocks. Eq. 4 actually shows a classical result, which is that light scattering can only provide a combination between some moments for the aggregation number distribution, and not the moments themselves. For large values of $\overline{n}$, the second term of the right hand side in Eq. 4 becomes negligible and the apparent molecular weight can be rewritten :

$$M_{w,\text{app}} = \left( \overline{n^2} / \overline{n} \right) m_n^* \tag{5}$$

Eq. 5 is probably more appropriate than the one generally used in the literature , and which expresses $M_{w,\text{app}}$ as the weighted sum of the weight-average molecular weights [12,17,18]. By choosing for building block a unit composed by one nanoparticle and two polymers (as we assume it is the case at $X = X_P$), the aggregation number featuring in the expression of $M_{w,\text{app}}$ (Eq. 5 and 6) coincides with $n^{nano}$. Using the values in Table I for the single components and a polydispersity of 1.6, the number-average molecular weight of the building blocks $m_n^*$ and the aggregation number $\overline{n^{nano^2}} / \overline{n^{nano}}$ can be estimated. We find that $\overline{n^{nano^2}} / \overline{n^{nano}}$ is of the order of $30 - 60$, whereas the number of polymers is about twice this value, between 60 and 120 (see Table I). These data are important since they provide a picture of the microstructure of the mixed aggregates. Note that under the condition of electroneutrality for the mixed aggregates, the number of two polymers per nanoparticle in the complexes sets up the structural charge for the yttrium hydroxyacetate about 140 (i.e. twice the degree of polymerization of the polyelectrolyte block). This suggests that the structural charge of the yttrium hydroxyacetate nanoparticles might be larger than for surfactant micelles. To have more accurate estimates of the core dimension and distribution function, scattering experiments at higher q ($> 10^{-3}$ Å$^{-1}$) are necessary.

# IV – Conclusions

In this paper we have shown the analogy between surfactant micelles and nanoparticles in the process of complexation with polyelectrolyte/neutral block copolymers. To demonstrate this analogy, we have used the same block copolymers for the two approaches. The copolymer is poly(sodium acrylate)-*b*-poly(acrylamide) with molecular weights 5000 and 60000 g mol$^{-1}$ and it is obtained by controlled radical polymerization in solution. The yttrium hydroxyacetate particles were chosen primarily as a model system for inorganic colloids, and also because of potential applications in the fields of ceramics and opto-electronic materials. Mixed aggregates are forming spontaneously by mixing dilute solutions of polymers and of nanoparticles. Inspired by the results obtained with micelles, the mixed colloids are also called "supermicellar" aggregates. In our previous reports, we used the equivalent terminology of colloidal complexes. Light scattering experiments is used quantitatively to determine the aggregation numbers of the "supermicelles". We have found that the aggregation numbers for the nanoparticles are in the range 30 – 60, depending on the mixing conditions, whereas the numbers of polymers is typically twice these values. As a comparison, in surfactant-polymer mixtures, aggregation numbers in the "supermicellar" aggregates were estimated of the order of hundreds, with typically one surfactant micelle per polymer. This suggests that the structural charge of the yttrium hydroxyacetate nanoparticles might be larger than for micelles. In a recent paper, we have also shown that the nanoparticle-polymer supermicelles exhibits a very goog long-term colloidal stability. Actually, it was found that they were by far more stable than the nanoparticles themselves [10]. The reason of this enhanced colloidal stability is the formation of a corona of the neutral blocks in the self-assembly process. The overall aggregates are neutral, or weakly charged and they interact with each others via soft steric interactions. The neutral corona has a typical thickness of about $20 - 30$ nm and it surrounds a core (of radius $\sim 10$ nm) containing the nanoparticles and the polyelectrolyte blocks. We suggest that the results shown here are quite general and that they could be easily extended to other types of nanoparticles, as for instance the class of the lanthanide hydroxyacetates.

**Acknowledgements** : We thank Yoann Lalatonne, J. Oberdisse, R. Schweins, A. Sehgal for many useful discussions, and Michel Rawiso for having pointed out to us the derivation of the apparent molecular weight for polydisperse associative colloids. Mathias Destarac from the Centre de Recherches d'Aubervilliers (Rhodia, France) is acknowledged for providing us the polymers. This research is supported by Rhodia and by the Centre de la Recherche Scientifique in France.

# References

[1] M. Barsoum, *Fundamentals of Ceramics* (McGraw-Hill International, 1997).

[2] Y.-M. Chiang, D. Birnie, W. D. Kingery, *Physical Ceramics - Principles for Ceramic Science and Engineering* (John Wiley anbd Sons, Inc., 1997).

[3] W. H. Rhodes, J. Am. Ceram. Soc. **64**, 19 (1981).

[4] C. He, Y. Guan, L. Yao, W. Cai, X. Li, Z. Yao, Mat. Res. Bull. **38**, 973 (2003).

[5] H. S. Roh, E. J. Kim, H. S. Kang, Y. C. Kang, H. D. Park, S. B. Park, Jpn. J. Apply. Phys. **42**, 2741 (2003).

[6] R. P. Rao, J. Electrochem. Soc., Vol. 143, No. 1, pp. 189-197 (1996). **143**, 189 (1996).

[7] W. Que, S. Buddhudu, Y. Zhou, Y. L. Lam, J. Zhou, Y. C. Chan, C. H. Kam, L. H. Gan, G. R. Deen, Mater. Sci. Eng. C **16, pp. 153-156 (2001).** 153 (2001).

[8] M. S. Tokumoto, S. Pulcinelli, C. V. Santilli, V. Briois, J. Phys. Chem. B **107**, 568 (2003).

[9] L. Kepinski, M. Zawadzki, W. Mista, Solid State Sci. (2004).

[10] K. Yokota, M. Morvan, J.-F. Berret, J. Oberdisse, Europhys. Lett., submitted (2004).

[11] K. Yokota, J.-F. Berret, B. Tolla, M. Morvan, US Patent No. RD 04004, serial number 60/540,430 (2004).

[12] K. Kataoka, H. Togawa, A. Harada, K. Yasugi, T. Matsumoto, S. Katayose, Macromolecules **29**, 8556 (1996).

[13] T. K. Bronich, A. V. Kabanov, V. A. Kabanov, K. Yui, A. Eisenberg, Macromolecules **30**, 3519 (1997).

[14] M. A. Cohen-Stuart, N. A. M. Besseling, R. G. Fokkink, Langmuir **14**, 6846 (1998).

[15] A. V. Kabanov, T. K. Bronich, V. A. Kabanov, K. Yu, A. Eisenberg, J. Am. Chem. Soc. **120**, 9941 (1998).

[16] T. K. Bronich, T. Cherry, S. Vinogradov, A. Eisenberg, V. A. Kabanov, A. V. Kabanov, Langmuir **14**, 6101 (1998).

[17] A. Harada, K. Kataoka, Macromolecules **31**, 288 (1998).

[18] A. Harada, K. Kataoka, Langmuir **15**, 4208 (1999).

[19] A. Harada, K. Kataoka, Science **283**, 65 (1999).

[20] T. K. Bronich, A. M. Popov, A. Eisenberg, V. A. Kabanov, A. V. Kabanov, Langmuir **16**, 481 (2000).

[21] K. Kataoka, A. Harada, Y. Nagasaki, Adv. Drug. Del. Rev. **47**, 113 (2001).

[22] P. Hervé, M. Destarac, J.-F. Berret, J. Lal, J. Oberdisse, I. Grillo, Europhys. Lett. **58**, 912 (2002).

[23] J.-F. Berret, G. Cristobal, P. Hervé, J. Oberdisse, I. Grillo, Eur. J. Phys. E **9**, 301 (2002).

[24] J.-F. Berret, P. Hervé, O. Aguerre-Chariol, J. Oberdisse, J. Phys. Chem. B **107**, 8111 (2003).

[25] C. Gérardin, N. Sanson, F. Bouyer, F. Fajula, J.-L. Puteaux, M. Joanicot, T. Chopin, Angew. Chem. Int. Ed. **42**, 3681 (2003).

[26] F. Bouyer, C. Gérardin, F. Fajula, J.-L. Puteaux, T. Chopin, Colloids Surf. A **217**, 179 (2003).

[27] J. H. Jeong, S. W. Kim, T. G. Park, Bioconjugate Chem. **14**, 473 (2003).

[28] J.-F. Berret, J. Oberdisse, Physica B **350**, 204 (2004).

[29] J.-F. Berret, B. Vigolo, R. Eng, P. Hervé, I. Grillo, L. Yang, Macromolecules **37**, 4922 (2004).

[30] S. v. d. Burgh, A. d. Keizer, M. A. Cohen-Stuart, Langmuir **20**, 1073 (2004).

[31] M. Castelnovo, Europhys. Lett. **62**, 841 (2003).

[32] M. Destarac, W. Bzducha, D. Taton, I. Gauthier-Gillaizeau, S. Z. Zard, Macromol. Rapid Commun. **23**, 1049 (2002).

[33] D. Taton, A.-Z. Wilczewska, M. Destarac, Macromol. Rapid Commun. **22**, 1497 (2001).

[34] L. Belloni, Colloids Surf. A **140**, 227 (1998).

[35] *Neutrons, X-rays and Light : Scattering Methods Applied to Soft Condensed Matter*, edited by P. Lindner, T. Zemb (Elsevier, Amsterdam, 2002).

[36] M. Rawiso, in *Diffusion des Neutrons aux Petits Angles*, edited by J.-P. Cotton, F. Nallet (EDP Sciences, Albé, France, 1999), Vol. 9, p. 147

[37] W. B. Russel, D. A. Saville, W. R. Schowalter, *Colloidal Dispersions* (Cambridge University Press, 1992).

[38] S. Förster, M. Zisenis, E. Wenz, M. Antonietti, J. Chem. Phys **104**, 9956 (1996).

[39] L. Willner, A. Pope, J. Allgaier, M. Monkenbusch, P. Lindner, D. Richter, Europhys. Lett. **51**, 628 (2000).

[40] S. Förster, N. Hermsdorf, C. Bo1ttcher, P. Lindner, Macromolecules **35**, 4096 (2002).

# Tables and Table Captions

| | dn/dc (cm$^3$ g$^{-1}$) | R$_H$ (nm) | R$_G$ (nm) | R$_G$/R$_H$ | M$_{w,app}$ (g mol$^{-1}$) | $\overline{n^{nano^2}}/\overline{n^{nano}}$ |
|---|---|---|---|---|---|---|
| block copolymer PANa(69)-b-PAM(840) | 0.159 | 8 | n.d. | n.d. | 68×10$^3$ | .. |
| nanoparticle Y(OH)$_{1.7}$(CH$_3$COO)$_{1.3}$ | 0.123 | ~ 2 | 1.65 ± 1 | ~ 0.8 | 27×10$^3$ | .. |
| mixed aggregate X = 0.2 | n.d. | 33.5 ± 1.5 | 19.0 ± 1.5 | 0.57 | 3.2×10$^6$ | 31 |
| mixed aggregate X = 0.25 | 0.150 | 42.0 ± 2 | 30.5 ± 1.5 | 0.73 | 5.8×10$^6$ | 57 |
| mixed aggregate X = 0.5$^{(*)}$ | n.d. | 28.5 ± 1.5 | 20.0 ± 1 | 0.70 | 2.5×10$^6$ | 29 |

**Table I** : Characteristic sizes, molecular weights and aggregation numbers of mixed aggregates made from nanoparticles and poly(acrylic acid)-*b*-poly(acrylamide) block copolymers. The hydrodynamic (R$_H$) and the gyration (R$_G$) radii, as well as the apparent molecular weight (M$_{w,app}$) of the colloidal complexes were determined using light scattering. The aggregation numbers $\overline{n^{nano^2}}/\overline{n^{nano}}$ are derived according to Eq. 5. (*) : For the series at X = 0.5, the aggregation number is calculated assuming that the mixed aggregates are at the preferred composition X$_P$ = 0.2.

# Figure Captions

**Figure 1 :** Representation of a colloidal complex formed by association between oppositely charged block copolymers and surfactants. The core is described as a complex coacervation micro-phase of micelles connected by the polyelectrolyte blocks. The corona is made of the neutral segments.

**Figure 2 :** Chemical structure of the poly(acrylic acid)-*b*-poly(acrylamide) diblock copolymers investigated in this work. The degrees of polymerisation of each block are m = 69 and n = 840, yielding a structure that is described in the text as PANa(69)-*b*-PAM(840). The polydispersity index for the diblock is 1.6.

**Figure 3 :** Evolution of (a) the Rayleigh ratio $\mathcal{R}$(q,**c**) and (b) the hydrodynamic radius $R_H$ as function of the charge ratio Z for mixed solutions of diblock copolymers PANa(69)-*b*-PAM(840) and oppositely charged surfactant DTAB (T = 25° C). The Rayleigh ratio is measured at q = 2.3×10⁻³ Å⁻¹ (θ = 90°). Each data points in the figures corresponds to a distinct solution, with total concentration c = 1 wt. %.

**Figure 4 :** Same as in Fig. 3 for solutions containing PANa(69)-*b*-PAM(840) copolymers and yttrium hydroxyacetate nanoparticles (c = 1 wt. %). The parameter X is the mixing ratio, i.e. the volume of yttrium-based sol relative to that of the polymer solution. The concentrations in nanoparticles and polymers in the mixed solutions are $c_{pol}$ = c/(1+X) and $c_{nano}$ = Xc/(1+X).

**Figure 5 :** Concentration dependence of the quadratic diffusion coefficient D(c) measured for mixed yttrium-polymer solutions prepared at c = 1 wt. % (X = 0.2, empty circles; X = 0.5, triangles) and at c = 4 wt. % (X = 0.25, closed circles). The extrapolation at zero concentration is the self-diffusion coefficient $D_0$, from which the hydrodynamic radius $R_H$ os derived. $R_H$ is found in the range 28 nm - 42 nm (see Table I for details).

**Figure 6 :** Zimm plot showing the evolution of the light scattering intensity measured for nanoparticle-polymer complexes as function of the wave-vector q (X = 0.2). Straight lines are calculated according to Eq. 1 using for fitting parameters $R_G = 19 \pm 1.5$ nm, $M_{w,app} = 3.2 \times 10^6$ g mol$^{-1}$.

**Figure 7 :** Same as in Fig. 6 for a series of solutions prepared at c = 4 wt. % (X = 0.25) and diluted thereafter down to c = 0.0165 wt. %. Straight lines are calculated according to Eq. 1 using $R_G = 30.5 \pm 15$ Å, $M_{w,app} = 5.8 \times 10^6$ g mol$^{-1}$ for fitting parameters.

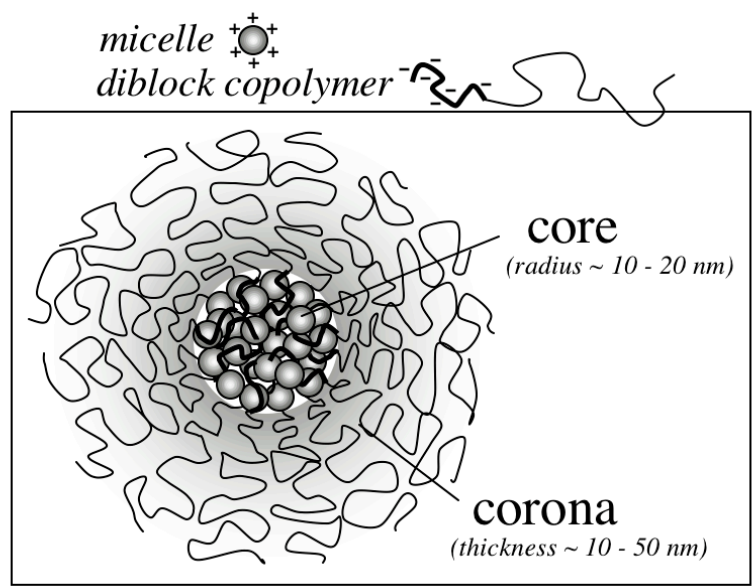

Figure 1

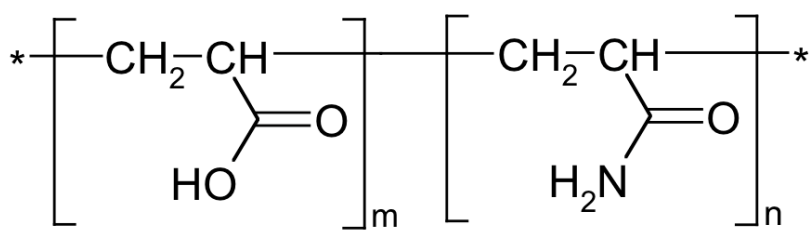

Figure 2

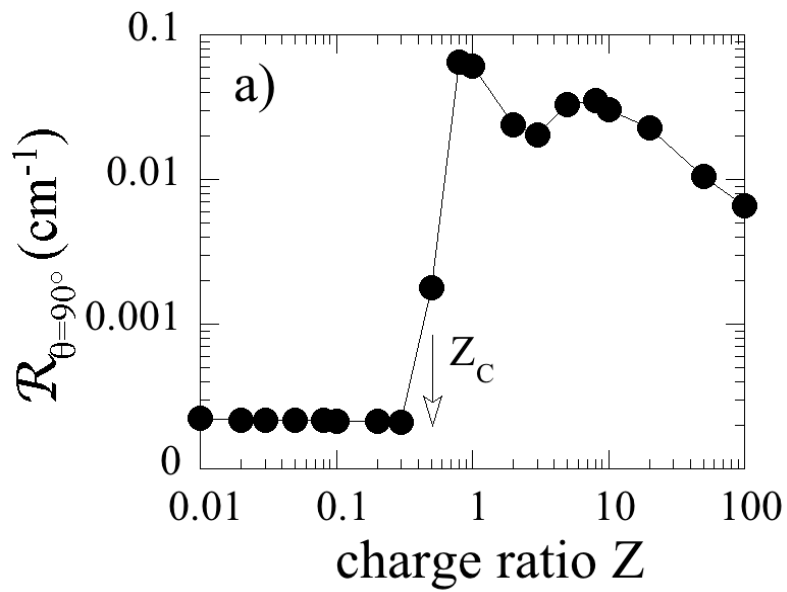

Figure 3a

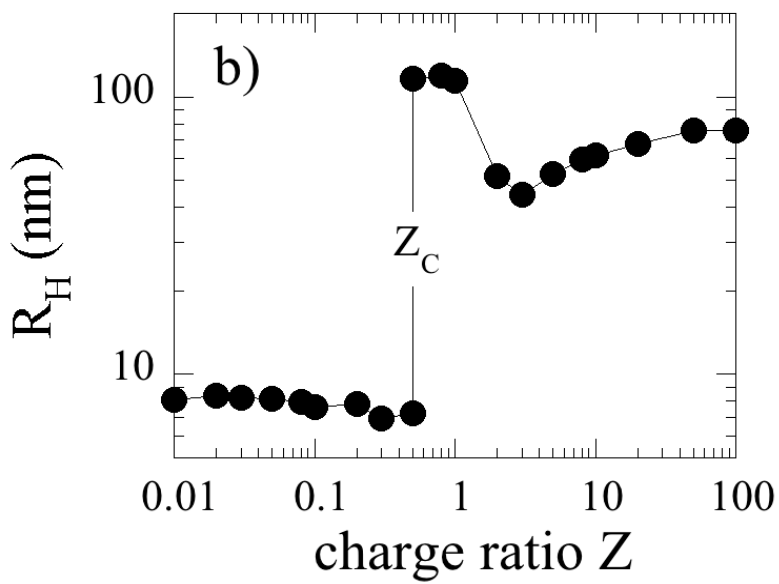

Figure 3b

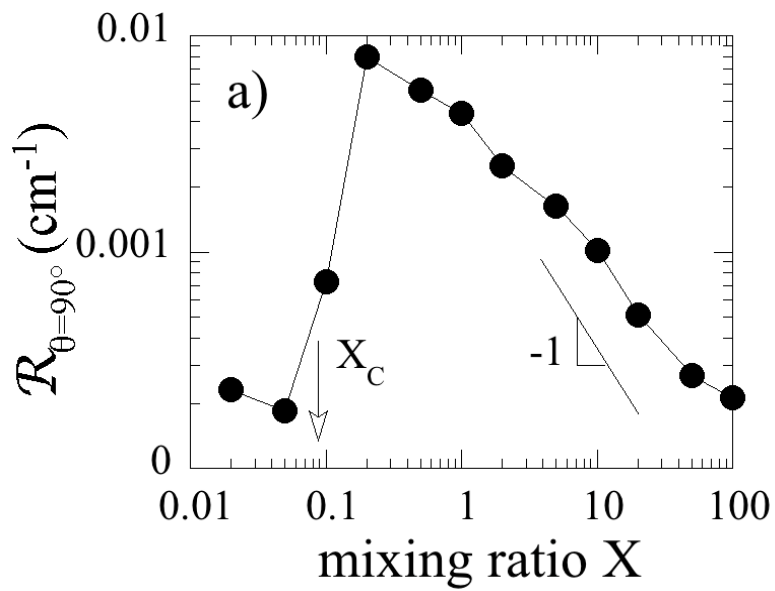

Figure 4a

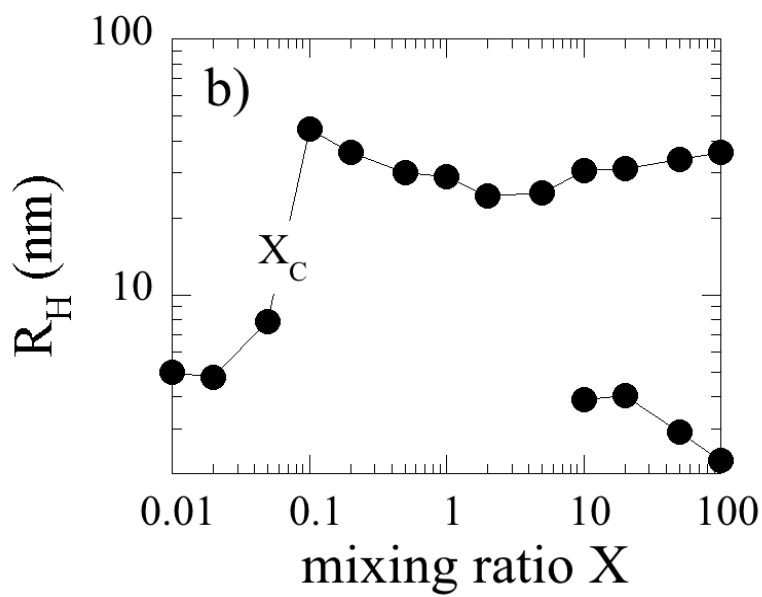

Figure 4b

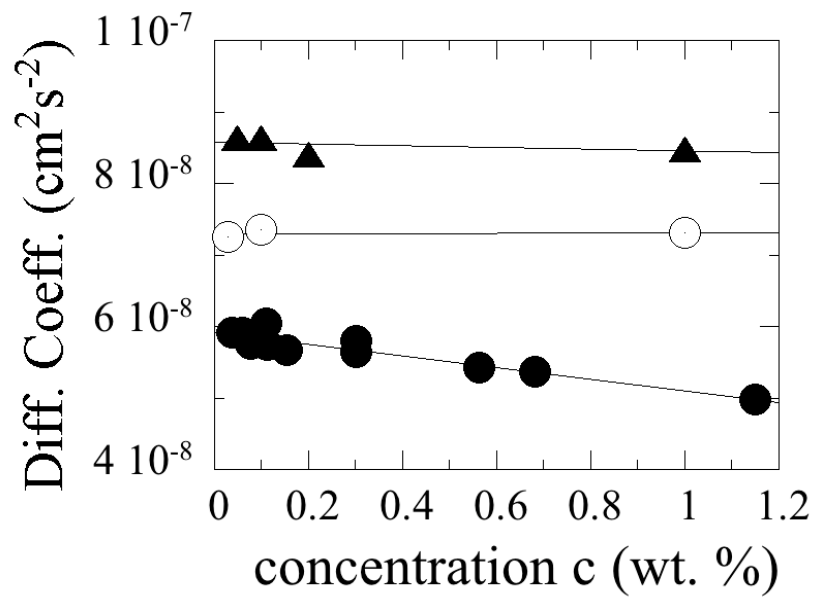

Figure 5

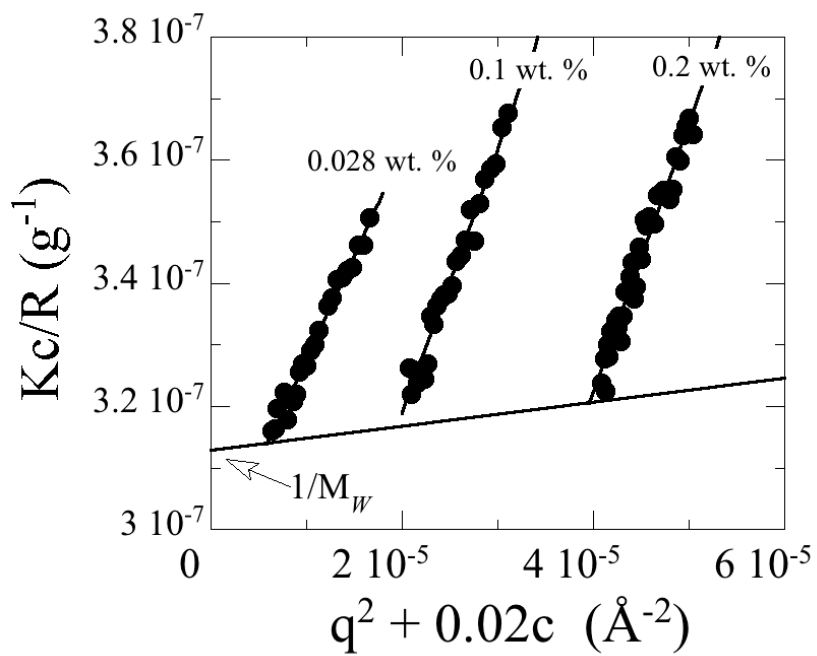

Figure 6

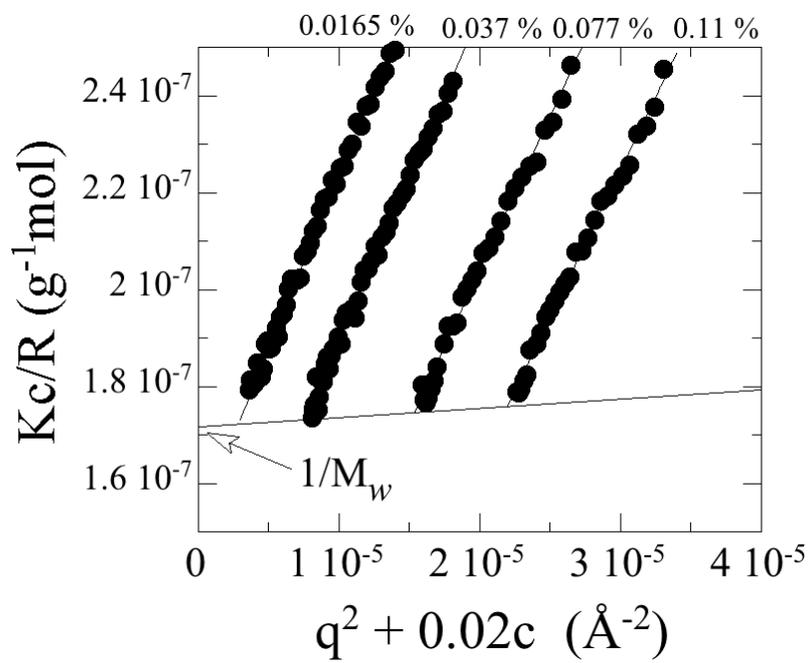

Figure 7